# GIS-AIDED SIMULATION OF SPATIAL DISTRIBUTION OF SOME POLLUTANTS AT "STOLBY" STATE RESERVATION

Marina G.Erunova, Krasnoyarsk State A & M University; *marina@icm.krasn.ru*

Anna A.Gosteva, Krasnoyarsk State Technical University, *anechka@torins.ru*

Michael G.Sadovsky, Krasnoyarsk State University, *msad@icm.krasn.ru*

Reserved territories seem to be the best reference sites of wildnature, where the long-term observations are carried out. Simulation model of spatially distributed processes of contamination of the state reservation is developed, and the dynamics of some pollutants is studied. An issue of the generalized evaluation of an ecological system status is discussed.

## I. Introduction

Rapid growth of human activities in natural resources consumption manifests in extensive human-induced ecological deterioration [1, 15], especially in the regions with severe climate and rather specialized environmental conditions [10, 13, 39]. The human affected ecosystems change significantly; it may be irreversible. For example, global pollution with non-ferrous and heavy metals leads to their accumulation at ecosystems. Further, they migrate over a web net [11, 21]. A researcher faces numerous problems in detail and comprehensive studying of the dynamics of a poisoned ecosystem, as well as the dynamics of various compartments of such ecosystems. Mathematical modelling is the powerful tool for such studies. Yet, the problems arises for a spatially distributed ecosystem. Reaction-diffusion equation system seems to be the most popular way to model such entities. Diffusion has nothing to do with real transfer of animals of any species [17, 18, 41, 42], even microorganisms [18, 43]; that is the basic discrepancy of this approach. Indeed, any migration tends to improve the life conditions and local habitat environment of an organism; such improvement resulted from a random and aimless walking is questionable.

Meanwhile, a comprehensive and reliable model methodology taking into account the non-random and targeted migrations of organisms is not developed yet, in detail [17, 41, 42]. Here, the power of simulation models is high enough. Simulation models may predict some nearer consequences rather precisely, at short-term scale. Long-term prognosis based on a simulation model is not reliable, in general. Thus, the verification of various simulation models of spatially distributed communities is of great importance; GIS-based digital models of territories involving the simulation of dynamics of various compartments of those sites may bring significant improvement the reliability of simulations (see, e.g., [9, 19, 30, 40]). In particular, GIS-based simulation models are the only tool providing a researcher with the method taking into account the effects of landscape. These latter are of great importance for numerous real ecological systems. Moreover, GIS-based model of a territory accounting both geographical, physical and environmental features of a site could be the best basis for the modelling of spatially distributed ecosystems [9].

A study of spatially distributed communities poses a question towards the connection of physical and geographical features of a territory and the biological features of that former. For example, it is a well known fact that various types of vegetation prefer to occupy different geographical areas. In particular, there is an acute question towards the relation of a climax (or stable) forest community and the altitude of the place of occupation, as well as the other geographical peculiarities (say, aspect, slope, etc.). Such landscape-scale models have been developed in attempt to predict a pattern of vegetation or other natural resources [6, 9, 12, 25, 33, 35, 56].

Global pollutant dispersion is the acute problem. Experts in various fields implement diverse approaches, methods, techniques and tools to address the problem; mathematical modelling is among them. Both analytical and simulation models are used here; besides, a researcher faces the problem of interpolation of the modelling results over a space. Here GIS addresses the problem best of all. The studies of the distribution of pollutants over the territories are carried out rather widely. Basically, such studies are targeted towards economy issues, rather than the environmental ones (see, e.g. [5, 20, 47]). The studies of protected areas/reservations with the help of GIS tools are not too convenient; some results in that direction see in [26, 34]. The studies of fully preserved territories located at Russia, in Central Siberia are presented by N.Prechtel [37, 38].

There exists a set of reserved territories (*zapovednik*) at Russia, founded for the purposes of preservation, research, monitoring and evaluation of the state-of-the-art of human impact on a wild nature. Such sites are the entities where no consumption of natural resources takes place, and the impact is eliminated to the highest possible degree. The sites are protected by law; special federal agency provides management, protection, control and monitoring at them. Since the reserved sites are the entities with maximally eliminated human impact, one can consider them as a reference natural ecological system. Such ecosystem allows a scientific forecasting of a natural environment evolution. The researchers refer to the ecosystem state at the sites, when studying an impact of human activity on a wildnature, and verify the methodology of the assessment of that latter.

Abundant data are gathered at such preserved sites during the period of their activity, at Russia, while an access to this information is impeded. To overcome this discrepancy, one should implement a systemic solution for storage, systematization and data processing of that information. Keeping in mind a growth of human impact, one sees a great importance of such solution. Modern computers powered by the up-to-date computational technologies allow development of the software tools for the investigations of ecosystems exposed to man-made change of environment. All the data collected in the reservations are spatially distributed, what makes a geographical information system (GIS) the adequate tool for storage, systematization and processing of those data. Public access to the collected data, as well as to other knowledge concerning the environmental situation is also of great importance. Modern GIS solutions provide both experts and public with the powerful tools for data representation, data visualization, remote processing and knowledge retrieval. Currently, Internet is the most efficient medium for such information access and exchange [6]. Basically new solutions based on Internet bring a closer cooperation of all entities interested in the problem. Such solutions improve and enforce the ecological and environmental education of residents worldwide. GIS atlas seems to be the most efficient method for presentation and processing of the monitored data of the natural processes run at a site; the efficiency of the atlas grows up due to implementation of Internet resources. Currently, there are few atlases of reserved sites available via Internet. Meanwhile, these atlases provide a consumer with visual information on the site, only.

This study is aimed to seek, figure out and clarify the statistical relationships between spatially-interpolated resource variables (vegetation type, altitude, soil type, river network, relief) and the distribution of some global pollutants, in the "Stolby" reservation (*zapovednik*). Main tasks of the paper are the followings:
- to reveal the statistical relationship between a forest type and some physical and geographical features of the landscape;
- to reveal the statistical relationship between a soil and some physical and geographical features of the landscape;
- to reveal the statistical relationship between a forest type and relevant soil, with respect to the features of landscape;
- to reveal the statistical relationship between the individual pollutant contamination and physical and geographical features of *zapovednik*, with respect to three biota compartments: soil, substrate and vegetation, and precipitations.

Here we also present the preliminary results of implementation of special ecological internet GIS-atlas of "Stolby" *zapovednik*. The atlas presented here differs from the existing ones by implementation of specific information analysis system. The system provides a researcher with a kit of various tools for information processing and analysis. User's manual of standardized techniques for data input, storage, treatment and visualization of diverse spatial data is provided with the atlas, as well.

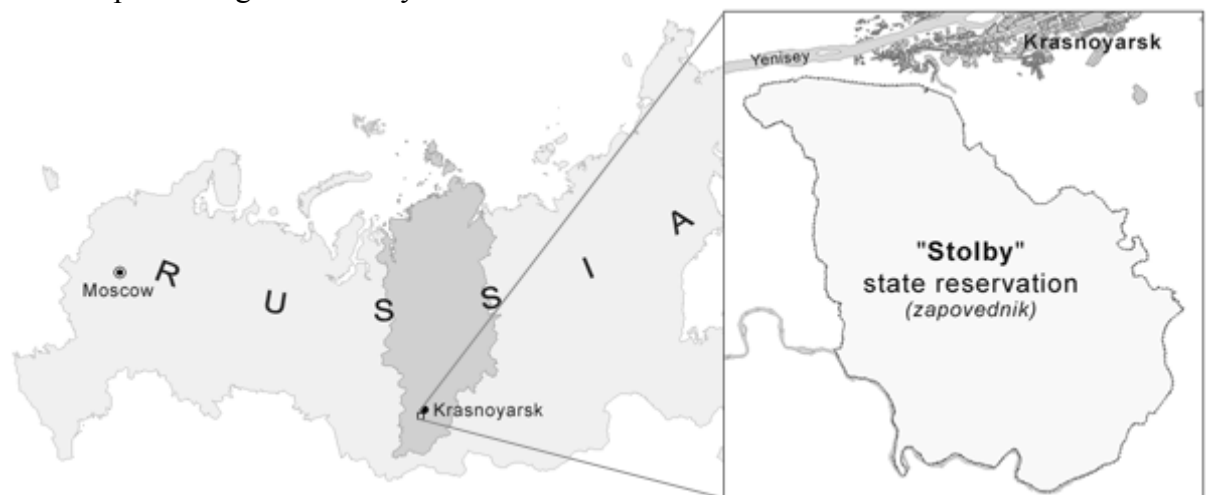

Fig. 1. Schematic location of "Stolby" State Reservation.

## II. Materials and methods

### 2.1. Study area

"Stolby" state reservation is located next to the large industrial city of Krasnoyarsk, on the right bank of the river of Yenisei (see Fig.1.). Its total area is 472 square kilometres; it is bordered by two tributaries of Yenisei, and occupies the mountain ridge system of East Sayan. The altitude within the site varies from 200 to 800 meters. *Zapovednik* was established in 1925; the regular observations had taken start at 1935. The site is unique from the point of view of a study of environmental processes including those resulted from human impact. Currently, the monitoring of man-caused pollution is heavily studied at the *zapovednik*, since that latter is a popular recreational place of the residents. The man-caused impact was restricted with a resident recreation. The data collected during that period allow well-grounded and profound investigation of the dynamics of the processes occurred in the ecological system of the site [29]. To collect the data, 48 observatory stations were grounded at the *zapovednik* territory. Each station provides four samples: the sample of soil, the sample of soil substrate, the sample of vegetation (pine needles, in our case), and the precipitation. The data were collected from the soil layer not exceeding 10 cm in depth. It should be mentioned, that the precipitations have been gathered in winter, only (thus, snow samples were collected). Such pattern of samplings results from a prevalence in wind, which blows from the city towards the site direction in winter, primarily.

### 2.2. Data collection – sampling description

To carry out the work, we developed the GIS-database incorporating the following data: relief data, soil type distribution data, river network data, road network data and the detailed forest estimation data. The forest estimation database specifies forty data fields. All these databases were converted into the digital form from paper records.

*Zapovedniks* are the reference points in the global monitoring. Monitoring is carried out at the reservation through the collection of the up-to-date data on the pollution level at the site. There is a network of stationary observatories located with respect to physical and geographical features of region, e.g. altitude, aspects, slopes, air currents, etc. The location of the observatories is determined by the administration of the reservation according to the corresponding legal and scientific issues. The observatories collect atmospheric precipitates, samples of forest substrate, of soil and vegetation.

They traced the concentrations of *Zn, Pb, Ni, Hg, F, As, Cd, Cu* and *Mn* at soil, for pollutant database. *Zn, Pb, Ni, Hg, F, As, Cd* and *Cu* have been traced for the forest substrate database. Data on the pollutant concentration in vegetation were collected through the sampling of pine needles; *Zn, F, Al, Cu, Mg* and *Mn* have been determined in it and supplied to the database. The data on the pollution in precipitation have been gathered through the snow sampling; *Al, F, Ni, Zn, Ca, Cl, Cu, K, Mg, Mn,* and *S* were traced in them [27]. Sample analysis for the above elements content was carried out at the Institute of biophysics of SD of RAS (at Krasnoyarsk) for *Ca, Mg, Fe, Cu, Mn, Zn,* and *Co* (by atomic absorption method). Sulphur was determined through titrimetry. *Al, Pb, Ni,* and *Cr* were determined by emission spectrometry. *K* and *Na* were detected by flame photometry, and fluorine was determined by toxicology test [7, 23, 36].

### 2.3. Complex estimation of pollution

It is a common practice that the results of monitoring of global pollutants are shown in generalized index. In general, such generalization yields a loss of some (maybe, important) details. To address the problem, we developed GIS-based model that produces such index. The index is provided by the formula:

$$K_c = \frac{C}{C_b}, \qquad (1)$$

where $K_c$ is the concentration factor of a chemical element, $C$ is the real content of a microelement, and $C_b$ is the background content of the microelement. $K_c$ indicates an excess power of the element content over an averaged background level of that former determined over a territory. The reference sites should be chosen in a manner to minimize an influence of a pollutant. Quite often, compartments of an ecosystem are poisoned with several elements. Here the total contamination index could be derived representing an effect of a group of elements. This factor is determined for all elements detected within a sample, and for a site determined via geochemical sampling [24]. Thus, the total pollution index $Z_c$ is defined as follows:

$$Z_c = \sum_{j=1}^{n} K_c^{(j)} - (n-1), \qquad (2)$$

where $K_c^{(j)}$ is the concentration index of *j*-th pollutant, and *n* is the number of pollutants.

Basically, the methodology for development of generalized index is based on the idea of maximum permissible concentration (MPC). The point is that MPC methodology fails to take into account the specific features of a region, or a specificity of a pollutant impact on environment, etc. This discrepancy was broken through due to the model implementation, that changes a background level values of pollutants in compartments for clarkes. Besides, the background indices for fluorine and heavy metals are not available for the site. These data were simulated with the observations on these elements carried out at the central part of Eastern Sayan, since that latter is close to the *zapovednik* from the geomorphologic point of view [3, 49].

### 2.3. GIS-based method
#### 2.3.1. Software description

The digital model of zapovednik was developed with GeoDraw and GeoGraph software packages implemented at the Institute of Geography of RAS; OziExplorer package was used to convert the data from GPS. The PRJBld program was used to make up the relations between the rectangular and geographic coordinates with GeoDraw for Windows software package. ArcGIS software package supplied with Spatial Analyst and 3D Analyst modules was used to carry out the spatial data analysis. Format transformation of cartography data was executed with MapInfo Professional GIS software package. Finally, table data were processed with MS Excel, StatSoft Statistica and other packages; some special utilities have been developed in Delphi for data conversion.

#### 2.3.2. Digital model development methodology

The primary geographic data were taken from print maps. Geography map of 1 : 25000 scale, vegetation map of 1 : 25000 scale and soil map of 1 : 50000 scale have been used. A transformation of printed map into a digital one is conveyed in three steps. Scanning of printed matter is the first step; image process-ing with GeoDraw vectorization processor is the second step. Finally, a compilation of all the layers into a single map is the third step.

An accuracy of the digital model of *zapovednik* was justified over the vegetation map, since that latter was the "worst" one. This map contains the maximal number of polygons (more, than 2100). An accuracy error was obtained for each polygon. An averaged error of the entire map was provided by median of a sample, since that former is mostly stable against the overshoots $M_e = 7.77$.

*2.3.3. Spatial analysis*

Various techniques of the spatial analysis [48] were used for a study of geographical features of the site, including
- ✓ primarily spatial analysis (positioning, object detection via their attributes, computations involving geometry objects of higher level);
- ✓ measurements (linear features of polygons, shape measures, distances);
- ✓ classification (thematic cartography, neighbourhoods search, filtering, buffers, slopes and aspects);
- ✓ statistically defined surfaces (relief digital models, interpolations, grid models);
- ✓ spatial distributions (triangulations, lines and polygons distributions, orientation of linear and polygonal objects);
- ✓ overlays (selection, mapping of point, linear and polygonal objects).

The GIS-based analysis was implemented to investigate the dependence of a vegetation distribution on an altitude, the dependence of soil distribution against an altitude, and mutual distribution of soil and vegetation. All the results of the spatial analysis were visualized.

*2.3.4. Monitoring of pollutants*

An unbiased speculation on the environmental situation should be based on a study of pollutant distribution in each point together with the interaction of pollutants in each environmental compartment, as well as the percolation of specific pollutants through different compartments. The heavy metal concentrations, as well as some halogens concentrations were used as the input data for soil pollution maps to study the distribution of metals in various compartments of *zapovednik*. ArcGIS software was used to implement an electronic map of pollution distribution, accomplished with Spatial Analyst. Spatial interpolation of data was carried out with Inverse Distance Weighted (IDW). Girding was performed based on a grid size of $15 \times 15$ m$^2$ using all of the input points available with fixed weight of an observation point. Here the closer a point is to the centre of the cell being estimated, the more weight it has in the averaging process.

We have developed the maps of mutual distributions of heavy metals and fluorine in the compartments of *zapovednik* ecosystem. An estimation of chemical contamination of substrate, soil, winter precipitates, and needles was provided by the concentration factor for each pollutant individually, and by the total contamination factor determined for each compartment, as well. In general, microelement content does not exceed MPC at the reservation.

### III. Results

*3.1. Nodel of zapovednik*
*3.1.1. Digital model implementation*

Comprehensive studies of the dynamics of pollutants migrations over the compartments of an ecological system could be provided with the relevant mathematical model. Since the complicated spatial pattern of the ecosystem of "Stolby" *zapovednik*, the simulation model of spatial dynamics based on GIS solution has been developed. The simulation model consists of the digital layers, and the data from numerous databases attributed to the points in those layers. All the points in the layers are split into two categories; the former incorporates the observations positioned at the layers in the relevant location, and the latter involves the modelled data obtained due to interpolation, with special respect to relief peculiarities.

The following layers [14] have been digitized: altitude isolines, reservation borders, rock pillars, river network (both polygonal, and linear objects), springs, forest roads, walking paths, check-points, compartment lines, elevation points, forest estimation, soils, and others, fifteen layers in total. The developed digital model of *zapovednik* includes 57 layers, totally.

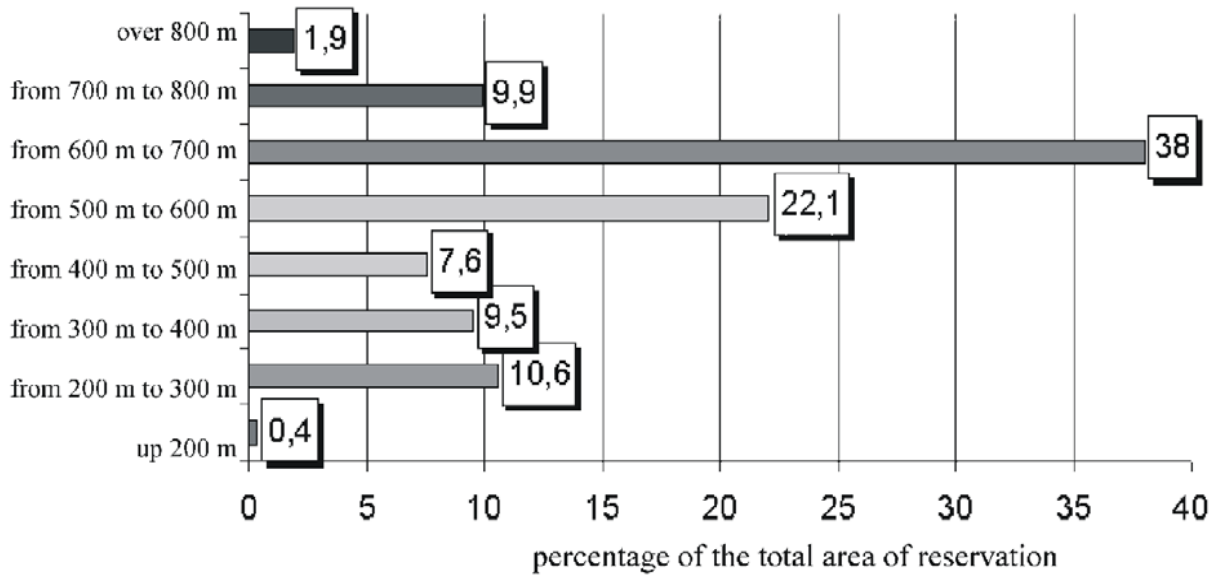

Fig. 2. Altitude distribution within the "Stolby" State reservation.

### 3.1.2. Spatial analysis of physics geography data

The analysis of distribution of forest types with respect to an altitude yields a prevalence of pine vegetation at almost any height (see Fig.2.). Few words should be said concerning the altitude pattern of the reservation. This latter is rather irregular, exhibiting significant variations in landscape features, say, slopes, mountain ridges, etc. Fig.2. shows a percentage of an area of the reservation in dependence on the height. Thus, Fig.3. shows the distribution of individual vegetation types with respect of a height indicates the absolute area measure occupied by a specific vegetation type at the given height. For example, pines occur mainly at the heath from 300 meters to 700 meters, occupying up to 25 km$^2$. The maximal area of pine occupation is observed at the altitude ranged from 500 meters to 520 meters. Similar pattern of a vegetation type distribution could be obtained for other species.

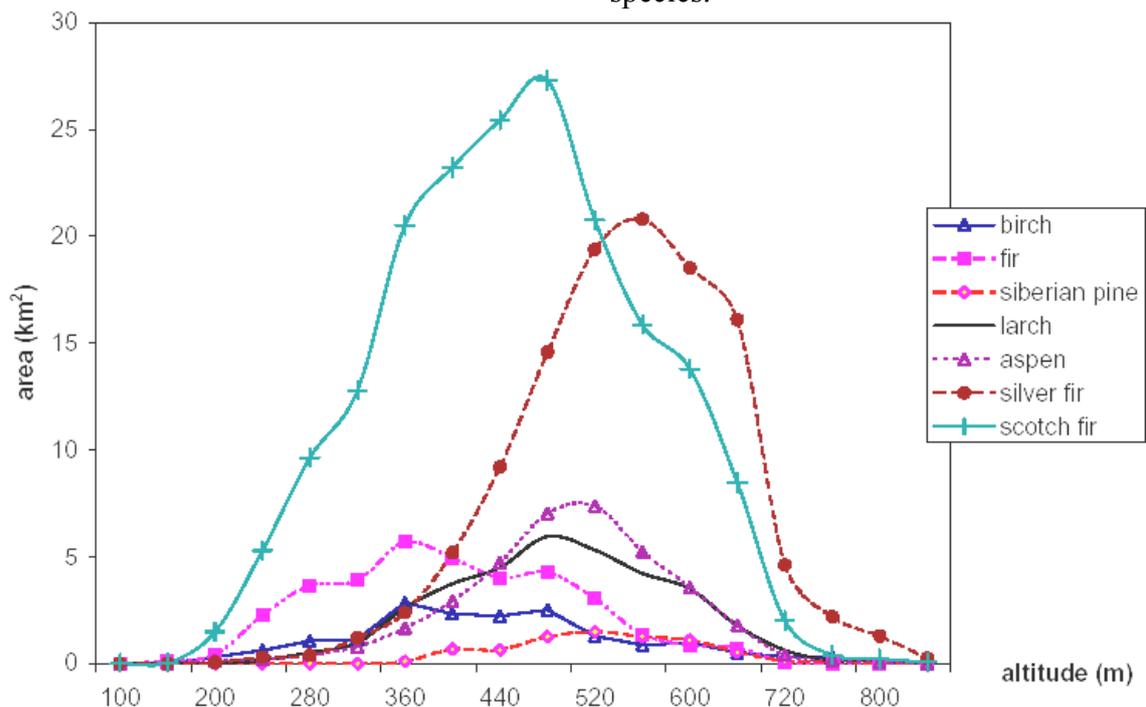

Fig. 3. Distribution of forest types over the altitude; the area of a type is calculated with altitude step equal to 40 meters.

An estimation of the contribution of each vegetation type for a specific height is shown in Fig.4. This figure represents the area of each vegetation type as a portion of the entire reservation area. This chart reveals the part of each vegetation type occupying the specific height. One easily sees the variation of the pattern of vegetation distribution with respect to the altitude. In contrast to Fig.3, this figure clearly presents the contribution of various forest species, within a specific height layer of the site. Besides, we have verified the vegetation map of "Stolby" state reservation, with particular respect to the succession of aspen for fir, for the last years. Large scale vegetation map (1 : 25 000) of the reservation presents the classification of flora, identifying 70 groups of forest types gathered into 21 series. This map mainly aims to figure out a spatial arrangement of groups and series of the types of biogeocenoses into basic structural entities: so called zone altitude complexes (ZAC).

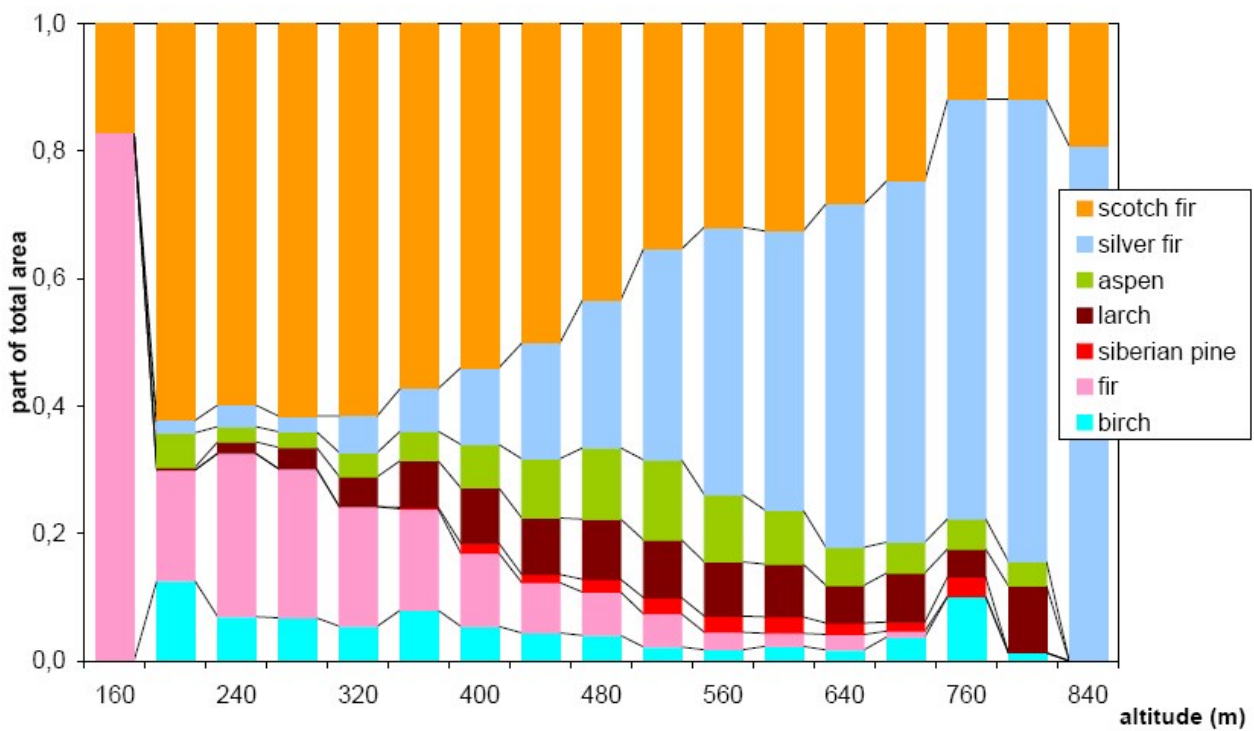

Fig. 4. Impact of each forest type for a given altitude; a part of area is calculated with altitude step equal to 40 meters.

### 3.2. Modelling of pollutant distribution

Distribution of chemical pollutants, as well as their migration pattern is affected with numerous complex factors. The behaviour of the entire complex of pollutants could hardly be described and present in simple figures; the point is that some pollutants are tended to a reasonably stable bunching, while others are not. Thus, one can quite precisely foresee the behaviour of some pollutants observing that latter of another. Here we figure out various groups of such combined bunching pollutants, with special attention to their mutual distribution and spatial dynamics. We shall start the analysis of results of modelling with the elucidation of the dynamics of the most motile element, and then consider similar dynamics of the elements with higher degree of immobilization. Next, the relations between various groups of pollutants, in connection to spatial features of the site, would be studied.

### 3.2.1. Spatial analysis of zinc distribution

To begin with, we start from the data on zinc distribution. Zinc is motile element at acid and low-acid soils, with moderate phytotoxicity. The growth of concentration of zinc yields a suppression of vegetation [11]. Zinc mainly is delivered to soil with industrial wastes from non-ferrous metallurgy, paintwork material industry, galvanism sewage and slugs of municipal sewage treatment facility.

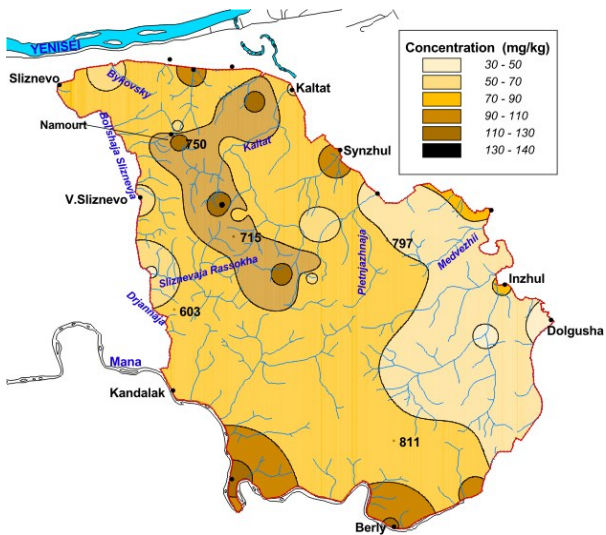

Fig. 5a. The pattern of zinc distribution at the substrate over *zapovednik*

The analysis of zinc distribution at the substrate (see Fig.5a.) shows that zinc content exceeds the background level, in general. There is a single site close to Synzhul checkpoint where the concentration is less than 40 mg/kg. Zinc contaminates the reservation, at most; its content ranges from 40 to 80 mg per kg, that twice exceeds the background level. The area located closely to Kaltat river, Sliznevaja river, Sliznevaja Rassokha river, and Drjannaja river, as well as close to Bykovsky spring are exposed to high zinc content (more than 80 mg/kg).

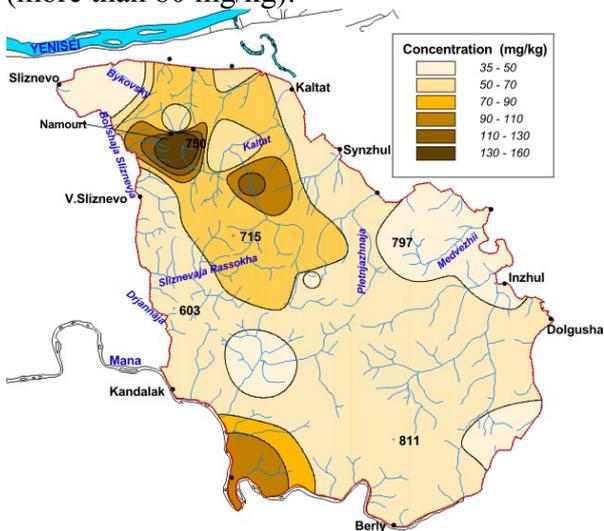

Fig.5b. The pattern of zinc distribution at soil over *zapovednik*

Zinc distribution in soil (see Fig.5b.) differs from that former in substrate. Basically, zinc content is close to the background level; this latter is equal to 51 mg/kg. Nevertheless, few spots of strong contamination are observed. In general, these spots coincide to the localities of the increased zinc content in the substrate. This fact probably follows from an active migration of zinc in the "substrate – soil" compartment, since one observes an acid medium there. Correlation coefficient of zinc content in substrate vs. that latter in soil is 0.63. Such correlation reveals vertical migration of zinc at the soil profile.

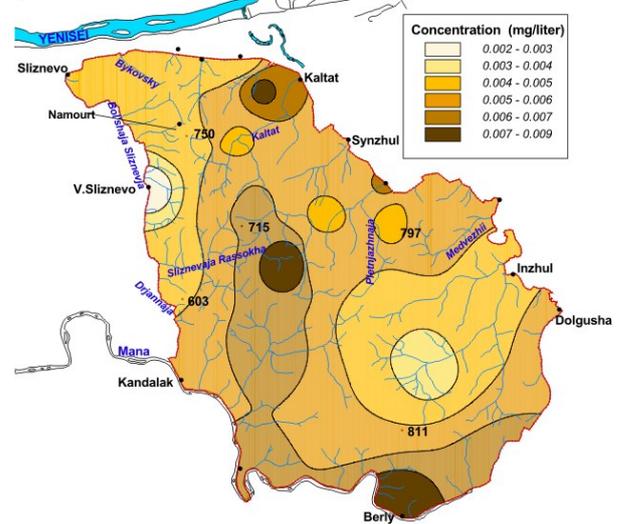

Fig.5c. The pattern of zinc distribution at winter precipitations over *zapovednik*

Increased zinc content in precipitates is observed at the watershed of Medvezhii spring and Pletnjazhnaja river, and at the area immediately neighbouring the suburb of the city of Kranoyarsk (see Fig.5c.). No congruence of the contamination spots is observed among needles, soil and substrate. Two regions exhibit the increased zinc content in vegetation (see Fig.5d.). The upper reaches of Namourt and Kaltat, and junction area of Sliznevaja Rassokha with Bol'shaja Sliznevja are these regions.

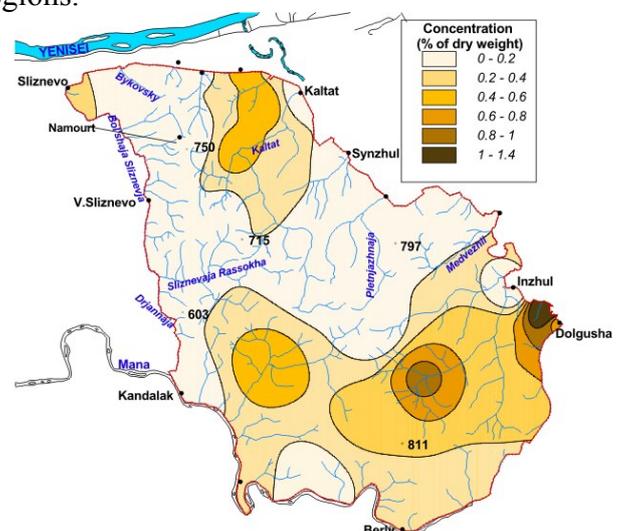

Fig.5d. The pattern of zinc distribution at vegetation over *zapovednik*

It should be noticed, that the areas of increased zinc content in vegetation are not related to the areas of the increased zinc content in substrate and upper soil level. Thus, zinc content in vegetation is related to the content of that former in rock, rather than to other compartments.

*3.2.2. Spatial analysis of fluorine distribution*

Fluorine is one of the most dangerous and motile elements in acid soils. Lethal atmospheric fluorine concentration for man is 0.0008 g per m$^3$. Fluorine excess in atmosphere results in extinguishing protective substances synthesis in plants [31]. *Zapovednik* is polluted by fluorine due to permanent operation of aluminium plant working for more than 30 years; its annual fluorine discharge in atmosphere exceeds 1000 tones; approximately 40 tones annually reaches *zapovednik*.

Increased fluorine content, under the moderately acid medium, is observed alongside the watershed of Kaydynski ridge. Eastern slopes exposed to the city of Krasnoyarsk are stronger poisoned. Similarly, the increased fluorine content in substrate is observed alongside the Kaydynski ridge watershed. The highest contamination with fluorine is observed at suburb area, where the content reaches 2.5 mg per kg, while the background level is equal to 1.15 mg per kg. Spatial analysis with maps overlay shows that the areas of increased fluorine content at soil and at substrate do no coincide. The correlation of fluorine content between these two compartments is 0.03. Fluorine migrates actively within soil, alongside the system of connected landscape patterns. The spots of increased soil contamination are tended to locate to the relief drops, reaching the spots with the lateral matter downstream. The maximal level of fluorine of 5 mg per kg is observed at the suburban territory, while the background level is equal to 1 mg per kg.

The increased fluorine content in precipitations is observed at the drops of the slopes, in peripheral area of *zapovednik*. Thematic map of its distribution looks similar to that latter for soil; increased fluorine content is observed at the accumulative parts of the landscape. Fluorine content analysis in vegetation reveals the growth of contamination in separate sites: in suburban area (Synzhul and Kandalak check-points); similar pattern of fluorine contamination is observed for substrate. The comparison of nidi of fluorine contamination in soil, substrate and needles one sees that the increased content of fluorine at needles is not related to the increased fluorine content in soil and substrate. The increased correlation between the fluorine contamination in all the compartments observed at Inzhul and Dolgusha check-points is eventual.

*3.2.3. Spatial analysis of lead distribution*

Lead is the most abundant pollutant affecting the health of population and environment worldwide. Lead pollution results from industrial and road traffic discharges thus yielding the significant pollution of suburb area. Lead affects human beings via respiration and food, being removed from organism rather slowly. In biosphere, lead is dispersed; it is not concentrated in a living matter. Potential intoxication of soils with lead grows up for more acid soils. Plants do not metabolize lead significantly. *Zapovednik* is contaminated with lead in part, with the accumulation of that latter in soil and substrate. The areas of Slizneva Rassokha, Dryannaya and Mana rivers exhibit the background lead content. An average lead content in soil and substrate (within the range of a double background level) is observed over the entire territory of *zapovednik*.

Comparative analysis of thematic maps of lead distribution shows that the area with increased lead content is greater, for surface, in comparison to the soil. Correlation coefficient of lead content between these two compartments is 0.51. Probably, this fact results from the increased human impact affecting the substrate. Lead poorly migrate into a soil, since the greater part is immobilized in the substrate. The increased lead content spot observed at the estuary of Bykovsky spring accompanied with the lesser lead content in the substrate probably results from the elimination of the element with superficial water. The area of watershed between the rivers of Laletina and Slizneva located close to Kaltat upper river is peculiar for the vertical redistribution of lead from the substrate into the soil. There is subacid soil there with very low superficial water discharge resulting is very low migration of lead. The lead content at the substrate observed within the suburban area does not exceed the background level, while it does at the soil exhibiting more than 15 mg per kg around the Laletina check-point close to Kitaiskaya Stenka rock. Background lead content here is equal to 10 mg per kg.

*3.2.4. Spatial analysis of mercury distribution*

Mercury poisons strongly all media (air, water and soil); it is contained in industrial wastes. Heat coal stations remain the basic source of the mercury pollution. It migrates into water with precipitations. Mercury is not accumulated in living organisms, migrating easily over ecosystems; it might be absorbed by clay and sludge. Neutral or alkalescent medium restricts the mercury migration. There are two nidi with maximal mercury content; the former is located at the watershed on Kaydynsky ridge, and the latter is located close to Bolshaya Sliznevaya river estuary. The least mercury content is observed at the southern end of *zapovednik*, close to Laletina and Narym check-points (that is the suburban territory), as well as at Kaltat upper river. The rest of territory exhibits an average mercury content (0.05 to 0.09 mg per kg).

Mercury distribution at soil differs from that one observed at the substrate, in general. The correlation between the mercury content observed at soil and substrate is 0.54. The main area of *zapovednik* exhibits a slightly increased mercury content (less than 0.05 mg per kg), while few spots of increased content of the element are observed. Background level of mercury at the soil is equal to 0.01 mg per kg. The highest mercury contamination is observed at the valley of Sukhoy Kaltat river and Mokry Kaltat river, and Inzhul river; the mercury content is equal to 0.06 mg per kg here. Other areas with increased mercury contamination are located over the entire territory of zapovednik, close to Kashtak check-point, and at the valleys of Bolshaya Slizneva and Mana rivers; here the content is about 0.05 mg per kg.

*3.2.5. Spatial analysis of arsenium distribution*

Arsenium is poorly migrating element; it contaminates portable water, air, vegetation, living organisms and soil. Usually, the arsenium contamination results from industrial pollution. Besides, the element is discharged into atmosphere due to natural geological processes, such as soil erosion, volcano activity and forest fires [8]. An increased arsenium content in soil and substrate is observed over the entire *zapovednik* territory. The worst situation is observed at the substrate. There are two nidi with maximal arsenium concentration; the former is at the suburb territory of *zapovedink* (Kaltat river estuary), the latter is the Magansk-Beret mountain pass. It should be said, that significant correlation is observed between soil and substrate content, both for increased content areas, and those being clear form the element.

*3.2.6. Spatial analysis of aluminium distribution*

Aluminium is rather toxic for plants. There are three regions at *zapovednik* with increased aluminium content. Vilistaya river area is the first region. Namurt and Kaltat upper rivers are the second region. Finally, the third region is the confluence place of Inzhul and Dolgy springs. In precipitations, aluminium is observed in increased values at three sites. These are the region of Kaltat check-point, the region of Dolgusha check-point, and Mana river valley where the accumulative landscape occurs. No aluminium contamination is observed at the central watershed ridge. The aluminium content at the needles strongly correlates to that one at winter precipitations.

*3.3. Modelling for complex estimation of pollution*

Objectively, an ecosystem status could be shown with the summary pollution index, which integrates the concentrations of various elements in that latter. Index $Z_c$ is the sum of excess of the factors of chemical compounds concentrations accumulated at sites of human activities (see formula (2)). The index varies from 13 to 145, for the substrate observed over the territory of *zapovednik*, with maximal values located at the suburban area. This area involves the region of Kitaisjaya Stenka rock, and the region of main road inside *zapovednik*; here the index is greater 128.

Index $Z_c$ varies from 7 to 35, for soils. The main body of the territory manifests the index value varying between 8 and 16. That falls within the proper range of the index. Pollution nidi at *zapovednik* manifesting the index $Z_c$ value between 16 and 32 are attracted to the suburb area, namely, to the main road inside the reservation and Verkhnyaya Sliznevo check-point, as well as to the mountain pass from Magansk to Beret. The suburb area is mostly poisoned with pollutants, manifesting $Z_c \leq 32$; this area is indicated as endangered one, with total index $32 \leq Z_c \leq 128$. Thus, one sees two regions with the highest pollution level; these are the region of Kitaiskaya Stenka rock and Kashtak check-point with Kuzmicheva lawn. Comparison of the maps of the individual elements distribution shows that four elements contribute most of all the contamination of soil; these are lead, zinc, mercury and nickel.

The areas of increased poisoning at the substrate are contaminated with mercury, nickel, zinc, fluorine and lead, while fluorine and mercury are the leading elements. Microelement content at the vegetation is usually determined by the biological features of that latter; also, bioaffinity and the content of the element at the root soil layer play the key role in the vegetation contamination. Generally, the integrated pollution index of needles varies from 1 to 2, that fall inside the permitted range. Considering the needles pollution, one clearly sees the contaminated area with $Z_c = 1.6$ around Krepost rock (rock Castle), which is close to the suburban area. The patterns of the total contamination in precipitations and in substrate are shown in Fig.6a, 6b.

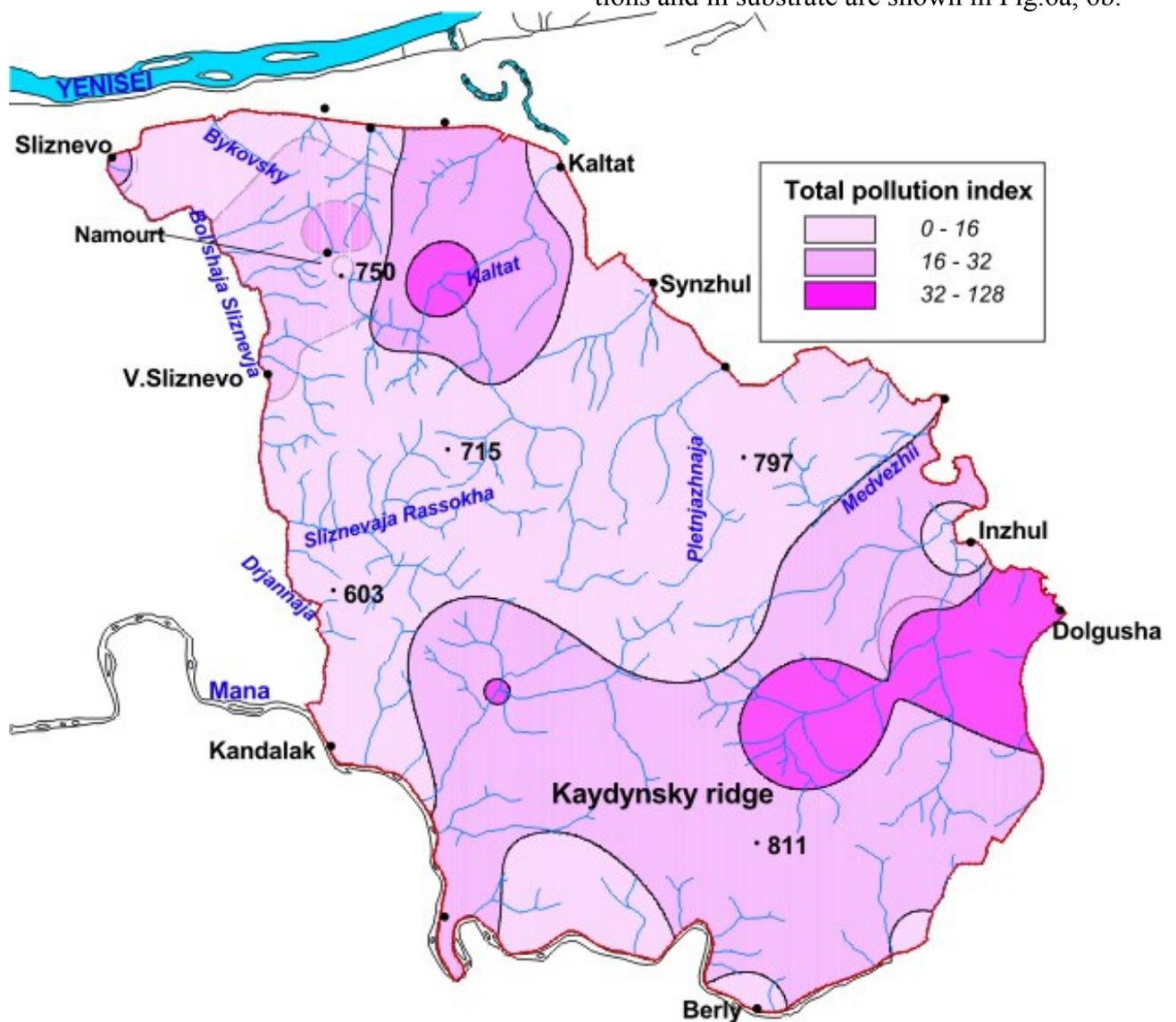

Fig.6a. Total contamination distribution map in winter precipitates

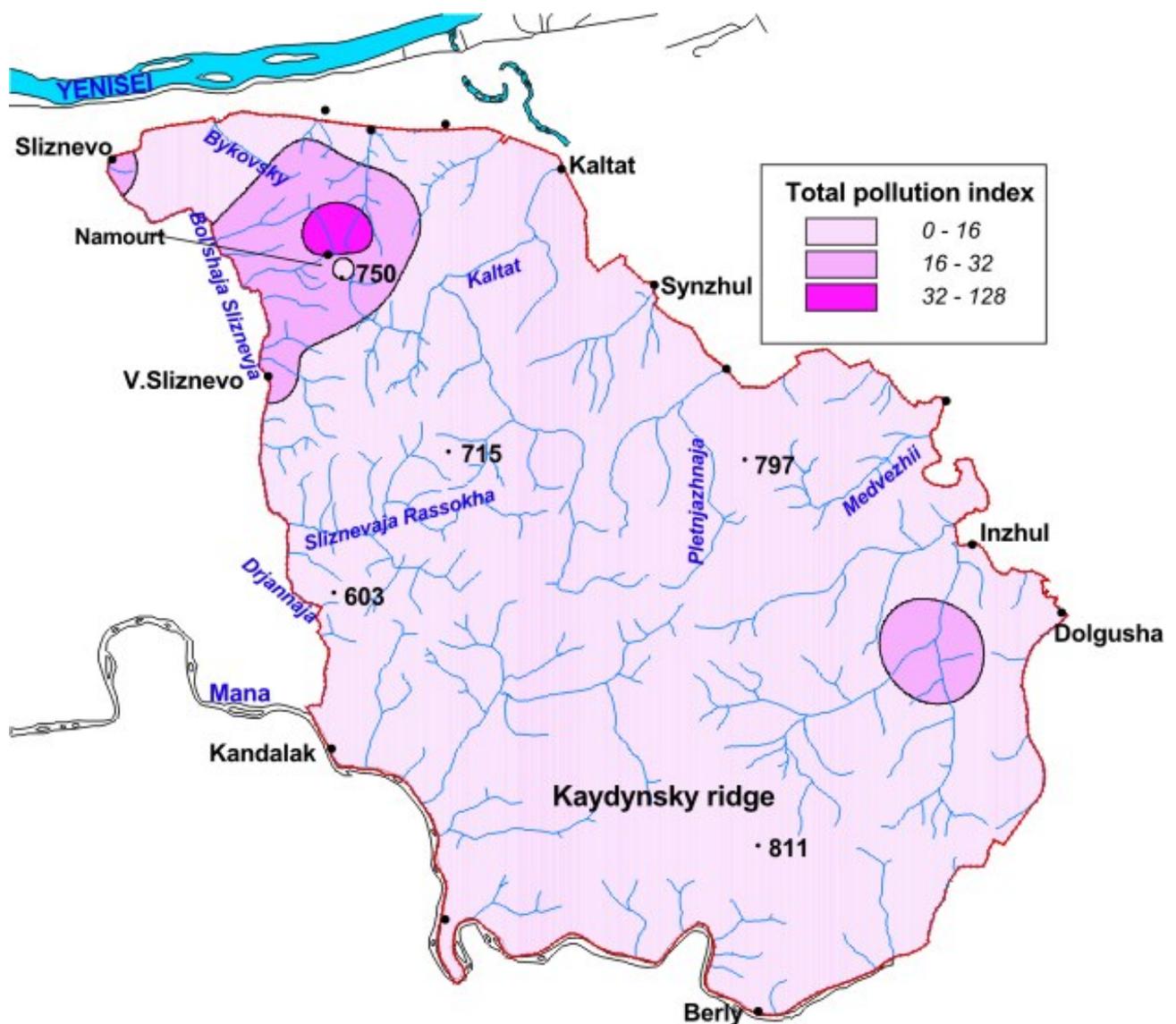

Fig.6b. Total contamination distribution map in substrate

The analysis of contamination in needles does not show an excess of pollution over the permitted level. This fact may follow from the increased resistance of conifers against the pollution, and criticality in vegetation growth occurs when pollutant level exceeds the background level in 3.4 to 4.5 times. Lichens are more sensitive, and criticality occurs at the 1.5 – 2.3 fold excess of the background level. Besides, they are very sensitive to air pollution. The studies of lichen flora have been carried out independently [28], it shows the highly correlated pattern of pollutant distribution. The generalized pollution distribution pattern is shown in Fig.7. The isolines of lichen viability exhibit rather tight correspondence to the pollutant distribution.

Lichen pollution indication methodology shows that the air pollution charge at the surface part of ecosystem falls within the 20 % to 40 % range of viability of lichens. It strongly corresponds to the data observed over the precipitations, needles and soil. The correlation results from the significant (up to 500 to 600 meters) higher location of the territory of *zapovednik* over the city, and the favourable wind occurrence. There are two anomalies at *zapovednik* located at suburban area and around Kaydynski mountain ridge; the data of the modelling and lichen observations correlates closely.

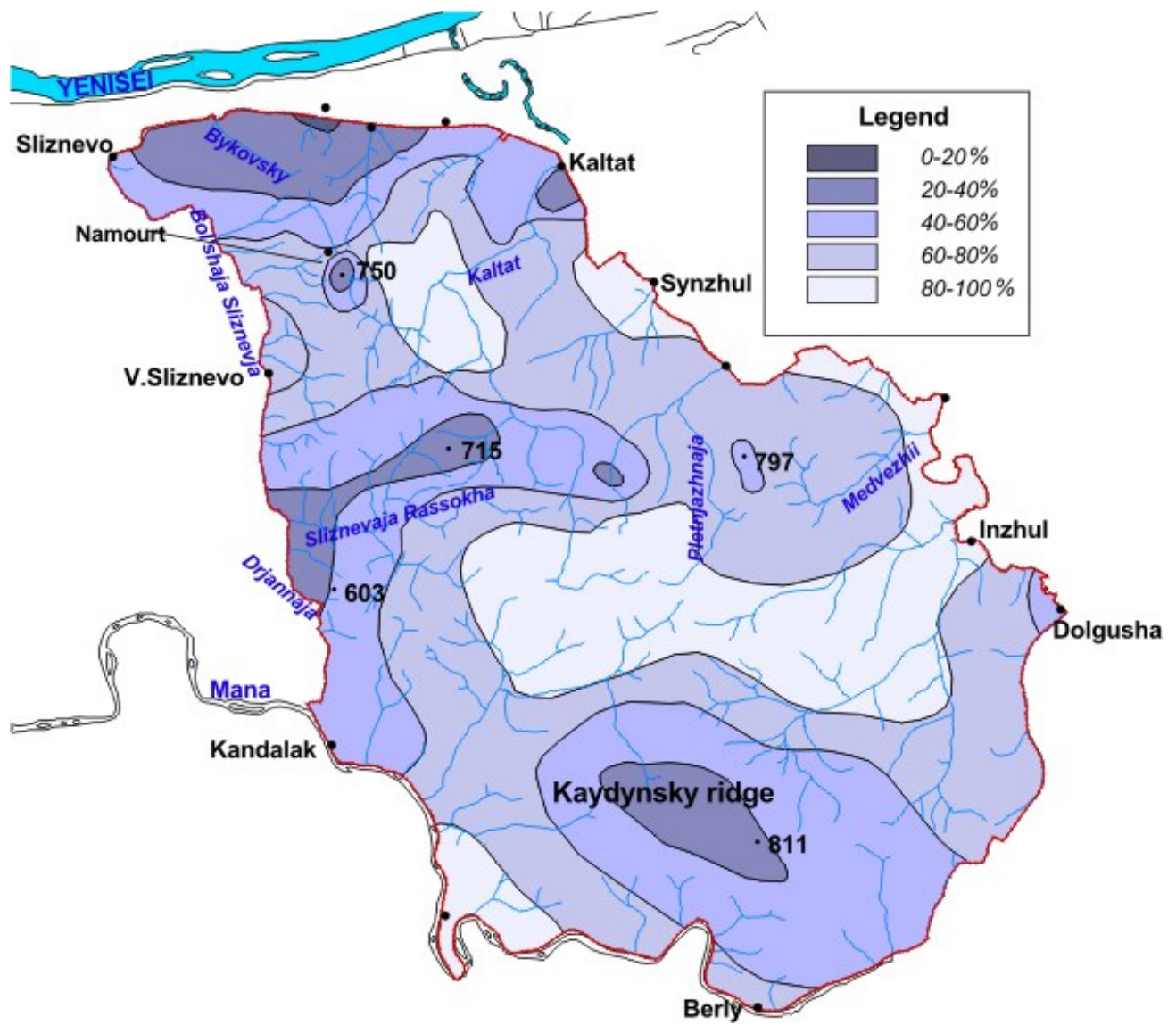

Fig.7. Lichen viability observed over the territory of *zapovednik*; the isolines identify the areas with permanent viability level.

**IV. Discussion**

Modelling of spatially distributed dynamics of various ecological systems still challenges the experts in various fields ranging from pure mathematics to specific issues in biology and computer science. The variety of approaches is great enough to meet all the aspects of these rather complicated studies. Clear, apparent and powerful modelling of the dynamics of compartments of spatially distributed ecological system is still doubtful, for many reasons. Here we present the results of the modelling of environmental processes running at the discrete and spatially extended ecological system of "Stolby" state reservation. We have implemented the simulation model of the site that figures out some typical steady regimes in the distribution of vegetation over the territory of *zapovednik*, and distribution of global pollutants affecting the ecosystem. GIS technologies and solutions are the key issues of the modelling implementation. Similar approach is present by Store and Jokimaki [45]. They discuss the model for habitat estimation, where this latter is affected by a number of factors. GIS implementation allows to figure out the effect of spatially distributed factors. Similarly, the modelling of generalized pollution pattern at *zapovednik* allows to evaluate the habitat conditions in that latter, in general. It should be said, that the studies of the environmental processes with the special emphasis to the reserved territories and reference sites with wild nature ecological pattern of habitation draw the attention of researchers (see, e.g., [37, 38]. These works present the modelling results of the environmental conditions at Katun State reservation located in Altai, where the impact of relief could hardly be taken into account but the GIS technologies and solutions.

It is a common fact, that environmental conditions impact the plant community and vegetation pattern of a (complex) ecological system, while the detailed pattern of such correspondence is not clear. The altitude of the vegetation place is the key factor here. We found rather stable and distinctive typical communities with prevalence of different forest species occupying the different altitude zones. Figs 3 and 4 show this relation between the altitude and forest species occupancy, in detail. Such complexes of dominating species are argued sometimes, while they are found by different researchers, in different climatic and environmental conditions; see, e.g., [37, 38] for Siberia, [9] for mountain China region. Quite detailed and smart study of such zone structure of forests is studied and discussed by B.Horsch [22].

The developed model reveals some statistical relations between all four compartments of the studied ecological system. Relief seems to be the leading factor here. Thus, soil types correlate quite closely to the altitude zones. Four types of soils are roughly identified at *zapovednik*; the types exhibit the prevalence with the altitude. This is not a point, since the types are mainly identified being based on the altitude zone structure. A relation between the class of soil and the altitude is less obvious. Eight classes of soil are determined at *zapovednik*; each class preferably occupies a specific altitude belt. A dispersion of the classes among themselves observed at *zapovednik* results from the highly jagged relief of the reservation.

On the contrary, one sees quite poor correlation between vegetation type, and soil class. One could expect to meet an increased correlation between these two compartments of the ecosystem, while there is observed rather poor correlation between these entities. The point is that soil formation is affected with a number of factors, including relief, rock, climate, season, vegetation, etc. All these factors force the relation simultaneously and, quite often, in opposite directions. Such complex interaction obturates discretion of the effect of a peculiar factor. More close study of the relations between soil class and vegetation reveals a decreased correlation among them. This fact means that the structure of forest associations fails to identify unambiguously the relevant class of soil. Similar observations are discussed in [22].

The efficiency of the approach to the modelling shown above should be used to study the dynamics of various pollutant distributions over an ecosystem. The dynamics of the transfer of such compounds could hardly be taken with the analytical modelling solely. The point is, that the pollutants are transferred both actively, by biotic components of an ecosystem, and passively, through the diffusion and other types of inactive transportation web. Such approach was used in [34], where the smart decision-making support system was implemented for the tasks of forest management and control. The GIS-based modelling is widely used for the studies of the dispersion of contaminations among the compartments of various ecosystems (see, e.g., [44]. The monitoring system supported with various Internet solutions becomes a new interactive scientific tool for a community of experts in various fields; see, e.g. *http://info.krasn.ru/stolby/* and *http://res.krasu.ru/ses/doc/1\_1.shtml* or [37, 38].

The results of the study presented here are mainly focused on the development of the simulation model describing the dynamics of vegetation and pollutant distribution for the purposes of environmental monitoring at "Stolby" reservation. The most up-to-date methods of cartography modelling and GIS-based analysis were used due to implement informational technologies of a complex analysis of spatial geographical data and environmental data. The methodology presented above is currently implemented for the studies of the "Tzentral'nosibirskii" State Biospheric Wildnature Reservation located at the southern part of Turukhansk region of Krasnoyarsk krai, and, partially, at Evenk autochthon area (Baikit region). The area of the reservation is

972017 hectares, exceeding more than 20 times the area of "Stolby" State Reservation. An efficiency of CIS technologies for the purposes of wildnature monitoring is evident; the technologies could be expanded for other territories, which meet a defence by society. A detail study due to GIS-technologies of the state and dynamics of natural processes at the "Tzentral'nosibirskii" State Biospheric Wildnature Reservation reveals the reference pattern of the dynamics of natural processes at the biosphere, since anthropogenic influence on the reservation is very low. The data gathered at the reservation could be used for estimation of the state of any reserved territory, both at Russia, and outside.